\documentclass[12pt]{article}
\usepackage{epsf}
\usepackage{amsmath, amssymb}
\usepackage{graphicx}

\begin{document}

\renewcommand{\thefootnote}{\fnsymbol{footnote}}
\setcounter{footnote}{0}
\begin{titlepage}

\def\thefootnote{\fnsymbol{footnote}}

\begin{center}

\hfill UT-11-17\\
\hfill May, 2011\\

\vskip .75in

{\Large \bf 
SUSY CP Problem in Gauge Mediation Model
}

\vskip .75in

{\large
Takeo Moroi and Norimi Yokozaki
}

\vskip 0.25in

{\em
Department of Physics, University of Tokyo,
Tokyo 113-0033, JAPAN}

\end{center}
\vskip .5in

\begin{abstract}

  SUSY CP problem in the gauge mediation supersymmetry breaking model
  is reconsidered.  We pay particular attention to two sources of CP
  violating phases whose effects were not seriously studied before;
  one is the effect of the breaking of the GUT relation among the
  gaugino masses due to the field responsible for the GUT symmetry
  breaking, and the other is the supergravity effect on the
  supersymmetry breaking parameters, in particular, on the bi-linear
  supersymmetry breaking Higgs mass term.  We show that both of them
  can induce too large electric dipole moments of electron, neutron,
  and so on, to be consistent with the experimental bounds.

\end{abstract}

\end{titlepage}

\renewcommand{\thepage}{\arabic{page}}
\setcounter{page}{1}
\renewcommand{\thefootnote}{\#\arabic{footnote}}
\setcounter{footnote}{0}

\section{Introduction}

Low-energy supersymmetry (SUSY), if it exists, affects various
low-energy phenomenology.  The most prominent place to look for the
effects of SUSY particles is the LHC experiment because it can provide
a direct confirmation of the existence of SUSY particles.  However,
precision experiments, in particular, those related to flavor and/or
CP violation, also put severe constraints on supersymmetric models.
The masses of superparticles are required to be larger than $10-100\
{\rm TeV}$ to avoid flavor and CP constraints if there is no mechanism
to suppress off-diagonal elements of the sfermion mass matrices and CP
violating phases \cite{Gabbiani:1996hi}.  Such a large value of the
superparticle masses are unacceptable from the naturalness point of
view.

One of the attractive mechanisms to avoid the SUSY flavor problem is
the gauge mediation \cite{Dine:1993yw, Dine:1994vc,
  Dine:1995ag}.\footnote
{For early attempts, see also \cite{Dine:1981za, Dimopoulos:1981au,
    Dine:1981gu, Nappi:1982hm, AlvarezGaume:1981wy}.}
In the gauge mediation model, sfermion masses are generated via the
interaction with gauge multiplet, so the degeneracy of the masses of
sfermions which have same gauge quantum numbers is guaranteed.  In
such a framework, however, it is still non-trivial to avoid the SUSY
CP problem.  This is because, even in the gauge mediation model, new
CP violating phases arise in operators which are flavor blind.  In
particular, relative phases among the gaugino masses, SUSY invariant
Higgs mass parameter, and the soft SUSY breaking bi-linear Higgs mass
(i.e., so-called $B_\mu$ parameter) do not vanish in general, which
often results in too large electric dipole moments (EDMs) of electron,
neutron, and so on \cite{Moroi:1998km}.  As in other cases, the SUSY
CP problem is  serious in the gauge mediation SUSY breaking
scenario.  So far, a possibility to solve the SUSY CP problem in the
gauge mediation model is thought to consider a model in which the
$B_\mu$ parameter vanishes at the messenger scale
\cite{Rattazzi:1996fb, Gabrielli:1997jp}.

In this Letter, we reconsider the SUSY CP problem.  We will
concentrate on the effects of CP violating phases which have not been
seriously considered before; we study two important effects which are
from (i) the breaking of the grand-unified-theory (GUT) relation among
the gaugino masses due to the GUT symmetry breaking field, and (ii)
the supergravity contribution to the $B_\mu$ parameter.  As we see in
the following, these possibly become sources of too large EDMs to be
consistent with the current constraints.  Thus, even if one adopts a
model with vanishing $B_\mu$ parameter at the messenger scale $M_{\rm
  mess}$, which has been regarded as a solution to the SUSY CP
problem, too large EDMs may be still induced.

\section{SUSY CP Problem in GMSB}

First, let us give an overview of the CP violating phases which are
relevant to our study.  In the present study, the most important CP
violating phases are relative phases among the gaugino masses $M_A$,
the SUSY invariant Higgs mass $\mu_H$, and the soft-SUSY breaking
Higgs mass parameter $B_\mu$.  (In the following, $A=1$, $2$, and $3$
correspond to $U(1)_{\rm Y}$, $SU(2)_{\rm L}$, and $SU(3)_{\rm C}$,
respectively.)

To see this, we denote the relevant part of the soft SUSY breaking
terms as
\begin{eqnarray}
  {\cal L}_{\rm soft} = - \frac{1}{2} \sum_A M_A \lambda_A \lambda_A 
  + B_\mu \mu_H H_u H_d + {\rm h.c.},
\end{eqnarray}
where $\lambda_A$ is the gaugino field, while $H_u$ and $H_d$ are up-
and down-type Higgs boson, respectively.  In the minimal
supersymmetric standard model (MSSM), the following phases are
invariant under the phase rotation of the MSSM fields and hence are
physical:\footnote
{The effects of the phases in the tri-linear coupling constants (i.e.,
  the so-called $A$ parameters) on EDMs do not have $\tan\beta$
  enhancement and are at most of the same order of those of $\phi_A$.
  In our numerical calculation, effects of the phases of
  $A$ parameters are properly taken into account.}
\begin{eqnarray}
  \phi_A = {\rm arg} (M_A B^*_\mu ).
\end{eqnarray}
If these phases are non-vanishing, they become the sources of EDMs
(and other CP violating quantities).  Thus, those phases are
stringently constrained although, in many models, they are expected to
be of $O(1)$.  In order to solve the SUSY CP problem, it is inevitable
to make all the phases of the gaugino masses equal; otherwise, it is
impossible to realize $\phi_1=\phi_2=\phi_3=0$.  In the simplest model
of GMSB, this is the case because the gaugino masses obey the GUT
relation.

One possibility of solving the SUSY CP problem is to adopt a model in
which the $B_\mu$ parameter vanishes at the messenger scale
\cite{Rattazzi:1996fb, Gabrielli:1997jp}.  In the gauge mediation
model, the scalar tri-linear coupling constants at the messenger scale
are expected to vanish (at least at the one-loop level).  Thus, we can
rotate away all the CP violating phases (other than the CKM phase)
from the MSSM Lagrangian if all the phases in the gaugino masses are
aligned.  Assuming that the low-energy effective theory below the
messenger scale is the MSSM, the $B_\mu$ parameter obeys the following
renormalization group equation
\begin{eqnarray}
  \frac{d B_\mu}{d \ln\mu} = 
  \frac{1}{8\pi^2} 
  \left[ 3 g_2^2 M_2 + g_1^2 M_1
    - 3 {\rm tr} (Y_U^\dagger A_U) - 3 {\rm tr} (Y_D^\dagger A_D)
    - {\rm tr} (Y_L^\dagger A_L)
  \right],
\end{eqnarray}
where $Y_U$, $Y_D$, and $Y_L$, are $3\times 3$ Yukawa matrices for
up-type quarks, down-type quarks, and leptons, respectively, while
$A_U$, $A_D$, and $A_L$, are corresponding tri-linear scalar coupling
matrices.  In addition, $g_2$ and $g_1$ are gauge coupling constants
for $SU(2)_{\rm L}$, and $U(1)_{\rm Y}$, respectively.  Then, even if
$B_\mu(M_{\rm mess})=0$, $B_\mu$ at the electroweak scale is generated
from the renormalization-group effect, which gives a viable Higgs
potential consistent with the electroweak symmetry breaking.

Another possibility is an accidental cancellation among the phases.
Even though such a cancellation requires a tuning of underlying
parameters, all the phases can be simultaneously made small if the
relation $\phi_1=\phi_2=\phi_3$ is realized.  This is the case if the
GUT relation among the gaugino masses holds.

In the following, we do not specify the mechanism to generate the
$\mu_H$ and $B_\mu$ parameters.  Our purpose is to study the effect of
the GUT symmetry breaking and the supergravity effect which have not
been studied in detail.  So we assume that $\phi_A$ vanish in the
limit that the above mentioned effects are neglected.

\section{Effect of GUT Relation Breaking}

First, we consider the CP violation from the breaking of the GUT
symmetry.  In simple models of gauge mediation which neglects the
effects of GUT symmetry breaking, it is guaranteed that
$\phi_1=\phi_2=\phi_3$.

We start our discussion with a general case; we parameterize the
superpotential of the messenger sector as
\begin{eqnarray}
  W = \sum_Q ( \lambda_{ij}^{(Q)} S + M_{ij}^{(Q)} ) \bar{Q}_i Q_j,
  \label{superpot}
\end{eqnarray}
where $Q$ and $\bar{Q}$ denote vector-like chiral multiplets which are
in irreducible representations of the standard-model gauge group
$G_{\rm SM}=SU(3)_{\rm C}\times SU(2)_{\rm L}\times U(1)_{\rm Y}$.
(The gauge indices are implicit.)  Here, $\lambda_{ij}$ is a coupling
constant while $M_{ij}$ a mass parameter.  In some of gauge mediation
models, in particular, those adopting the ISS mechanism of SUSY
breaking \cite{Intriligator:2006dd}, $M_{ij}$ is non-vanishing
\cite{Kitano:2010fa}, although there exist models with $M_{ij}=0$.
(For models with $M_{ij}\neq 0$, see also \cite{Murayama:2006yf,
  Murayama:2007fe}.)  As we will see, the existence of the mass
parameter significantly affects the aspect of CP violation in the SUSY
breaking parameters.  In general, there may exist several multiplets
in the same representation of $G_{\rm SM}$; those multiplets are
distinguished by the indices $i$ and $j$ which run $1-N^{(Q)}$, with
$N^{(Q)}$ being the number of $Q$ and $\bar{Q}$ pair.  Then, the
summation in Eq.\ \eqref{superpot} is over all the irreducible
representation of the $G_{\rm SM}$.  In addition, $S$ is the singlet
superfield whose $F$-component is non-vanishing; if there exists
several singlet fields, we take their linear combination such that
only $S$ acquires $F$-component VEV.  It should be understood that
$M^{(Q)}$ contains SUSY invariant mass of $Q$ and $\bar{Q}$ arising
from the VEV of singlet fields other than $S$.  

At the leading order of $F_S$, the gaugino mass is given by
\begin{eqnarray}
  m_A = \frac{g_A^2}{16\pi^2} \sum_Q b_A^{(Q)} \Lambda^{(Q)},
\end{eqnarray}
where $b_A^{(Q)}$ is the $\beta$-function coefficient of the gauge
coupling constant $g_A$ due to the chiral multiplet $Q$.  In addition,
$\Lambda^{(Q)}$ is the ratio between the SUSY breaking mass squared
and the supersymmetric mass of the multiplet $Q$ and is given by
\begin{eqnarray}
  \Lambda^{(Q)} \equiv 
  F_S \lambda^{(Q)}_{ij} 
  [\lambda^{(Q)} \langle S\rangle + M^{(Q)}]^{-1}_{ji},
\end{eqnarray}
with $[\lambda^{(Q)} \langle S\rangle + M^{(Q)}]^{-1}$ being the
inverse matrix of $\lambda^{(Q)}_{ij} \langle S\rangle +
M^{(Q)}_{ij}$.  

When there is no supersymmetric mass term (i.e., $M^{(Q)}=0$), we can
easily see that the phases of all the gaugino masses are aligned;
indeed, in such a case, $\Lambda^{(Q)}$ becomes $N^{(Q)}F_S/\langle
S\rangle$ and is independent of the parameters in superpotential.
Notice that this conclusion is independent of the assumption of GUT.

Even if there exists supersymmetric mass term, the phases of gaugino
masses become the same as long as the exact GUT relations hold among
$\lambda^{(Q)}_{ij}$ and $M^{(Q)}_{ij}$.  If $Q$ and $Q'$ are in a
same irreducible representation of GUT group $G_{\rm GUT}$, the
relations $\lambda^{(Q)}_{ij}=\lambda^{(Q')}_{ij}$ and
$M^{(Q)}_{ij}=M^{(Q')}_{ij}$ are realized with neglecting $G_{\rm
  GUT}$ breaking effects.  The above relations, however, break down
once we take account of the effect of $G_{\rm GUT}$ breaking.

It is obvious that the relation $\phi_1=\phi_2=\phi_3$ is not
guaranteed if the GUT relation
$\lambda^{(Q)}_{ij}=\lambda^{(Q')}_{ij}$ or
$M^{(Q)}_{ij}=M^{(Q')}_{ij}$ is somehow violated for $Q$ and $Q'$ from
the same multiplet of $G_{\rm GUT}$.  Violation may happen due to
operators including the field responsible for the GUT symmetry
breaking $G_{\rm GUT}\rightarrow G_{\rm SM}$.  (We denote such a field
$\Sigma$.)

We assume that all the higher-dimensional operators allowed by the
$G_{\rm GUT}$ exist with the cut-off scale of the order of the Planck
scale $M_{\rm Pl}\simeq 2.4\times 10^{18}\ {\rm GeV}$.  Then, their
effects are expected to be proportional to powers of
$\langle\Sigma\rangle/M_{\rm Pl}$, where $\langle\cdots\rangle$
denotes vacuum expectation value (VEV).  The VEV
$\langle\Sigma\rangle$ is expected to be of the order of the GUT scale
$M_{\rm GUT}\simeq 2\times 10^{16}\ {\rm GeV}$.  Importantly, the
higher-dimensional operators not only induce the breaking of the GUT
relations but also provide a new source of CP violation because the
coefficients of the higher-dimensional operators are complex in
general.  Even though the effect of the higher-dimensional operator is
expected to be suppressed by powers of $M_{\rm GUT}/M_{\rm Pl}\sim
10^{-2}$, it can be large enough to be constrained by present and
future experiments.  In particular, sizable EDMs can be generated, as
we see below.

The size of the breaking of the GUT relation is determined by the
structure of the superpotential, which depends on the properties of
the $G_{\rm GUT}$ breaking field $\Sigma$ and the vector-like
messenger multiplet.  In order to make our discussion concrete and
quantitative, we consider the simplest gauge mediation model based on
$SU(5)$ SUSY GUT; the messenger multiplet is assumed to consist of
${\bf 5}$ and ${\bf \bar{5}}$ representations of $SU(5)_{\rm GUT}$,
which we denote $\Psi$ and $\bar{\Psi}$, respectively.  Notice that
${\bf 5}\otimes {\bf \bar{5}}= {\bf 1}\oplus {\bf 24}$.  Thus, if the
GUT symmetry is broken by a field in ${\bf 24}$ representation, the
effect of GUT breaking can be of the order of $M_{\rm GUT}/M_{\rm
  Pl}$; otherwise, the effect is more suppressed.  Below the GUT
scale, the vector-like multiplets split into $\psi_d$, $\psi_l$ (which
are from $\Psi$), $\bar{\psi}_d$ and $\bar{\psi}_l$ (which are from
$\bar{\Psi}$), whose transformation properties under the
standard-model gauge group are $({\bf 3}, {\bf 1}, -1/3)$, $({\bf 1},
{\bf 2}, 1/2)$, $({\bf \bar{3}}, {\bf 1}, 1/3)$, and $({\bf 1}, {\bf
  2}, -1/2)$, respectively.

The first case is that $G_{\rm GUT}$ is broken by a field which
transforms as ${\bf 24}$ of $SU(5)_{\rm GUT}$, as in the minimal SUSY
GUT \cite{Dimopoulos:1981zb,Sakai:1981gr}.  We denote the GUT breaking
field as $\Sigma^\alpha_\beta$ (where $\alpha$ and $\beta$ are
$SU(5)_{\rm GUT}$ indices, which run $1-5$), whose vacuum-expectation
value is parameterized as
\begin{eqnarray}
  \langle \Sigma \rangle = \frac{v_{\bf 24}}{2\sqrt{15}}
  {\rm diag} (2,2,2,-3,-3).
\end{eqnarray}
Then the mass of $X$- and $Y$-bosons is given by
$m_{X,Y}=\sqrt{\frac{5}{6}}g_5 v_{\bf 24}$, where $g_5$ is the gauge
coupling constant at the GUT scale.  In this case, the effects of the
breaking of the GUT relation is proportional to $v_{\bf 24}/M_{\rm
  Pl}$:\footnote
{GUT symmetry breaking effect can be also on the mass terms, which may
  induce the EDMs.  The sizes of the EDMs are of the same order of
  those with the GUT symmetry breaking effect on the tri-linear term
  in the superpotential as far as the correction to the mass term is
  of the order of $\epsilon_{\bf 24}M$.}
\begin{eqnarray}
  W &=& \lambda_0 S \bar{\Psi}^\alpha \Psi_\alpha
  + \frac{\lambda_1}{M_{\rm Pl}} 
  S \bar{\Psi}^\alpha \Sigma^\beta_\alpha \Psi_\beta
  + M \bar{\Psi}^\alpha \Psi_\alpha
  \nonumber \\ &=&
  \lambda_0 \left( 1 + \frac{1}{\sqrt{15}}\epsilon_{\bf 24} \right )
  S \bar{\psi}_d \psi_d 
  + \lambda_0 \left( 1 - \frac{3}{2\sqrt{15}}\epsilon_{\bf 24} \right )
  S \bar{\psi}_l \psi_l + \cdots,
\end{eqnarray}
where 
\begin{eqnarray}
  \epsilon_{\bf 24} = \frac{\lambda_1 v_{\bf 24}}{\lambda_0 M_{\rm Pl}}.
\end{eqnarray}
Then, assuming $\lambda_0 \langle S\rangle\ll M$, the gaugino masses
are given by
\begin{eqnarray}
  M_A &=& \frac{g_A^2}{16\pi^2} 
  (1 + \kappa^{({\bf 24})}_A \epsilon_{\bf 24})
  \Lambda, \label{eq:24}
\end{eqnarray}
with $\kappa^{({\bf 24})}_1=-\frac{1}{2\sqrt{15}}$, $\kappa^{({\bf
    24})}_2=-\frac{3}{2\sqrt{15}}$, $\kappa^{({\bf
    24})}_3=\frac{1}{\sqrt{15}}$, and
\begin{eqnarray}
  \Lambda = \frac{\lambda_0 F_S}{\lambda_0 \langle S\rangle + M}.
\end{eqnarray}
Here, we only consider the leading-order contribution from $v_{\bf
  24}$.

In this framework, we calculate the EDMs of the electron, the neutron
and the mercury.  As we will see below, since the phase of $B_\mu$ is
not aligned to those of gaugino masses, the physical phase $\phi_A$
become non-negligible, resulting in sizable EDMs.  Here, we take
$|\epsilon_{\bf 24}|=10^{-2}$, and ${\rm arg}(\epsilon_{\bf
  24})=\frac{\pi}{2}$; the phase is chosen to maximize the
contribution to the EDMs.  With this choice of the parameters, the
physical phase $\phi_A$ of the order of $O(10^{-2})$ is induced, and
this phase is large enough to be constrained from the present EDM
experiments.  

\begin{figure}
  \centerline{\epsfysize=0.8\textheight\epsfbox{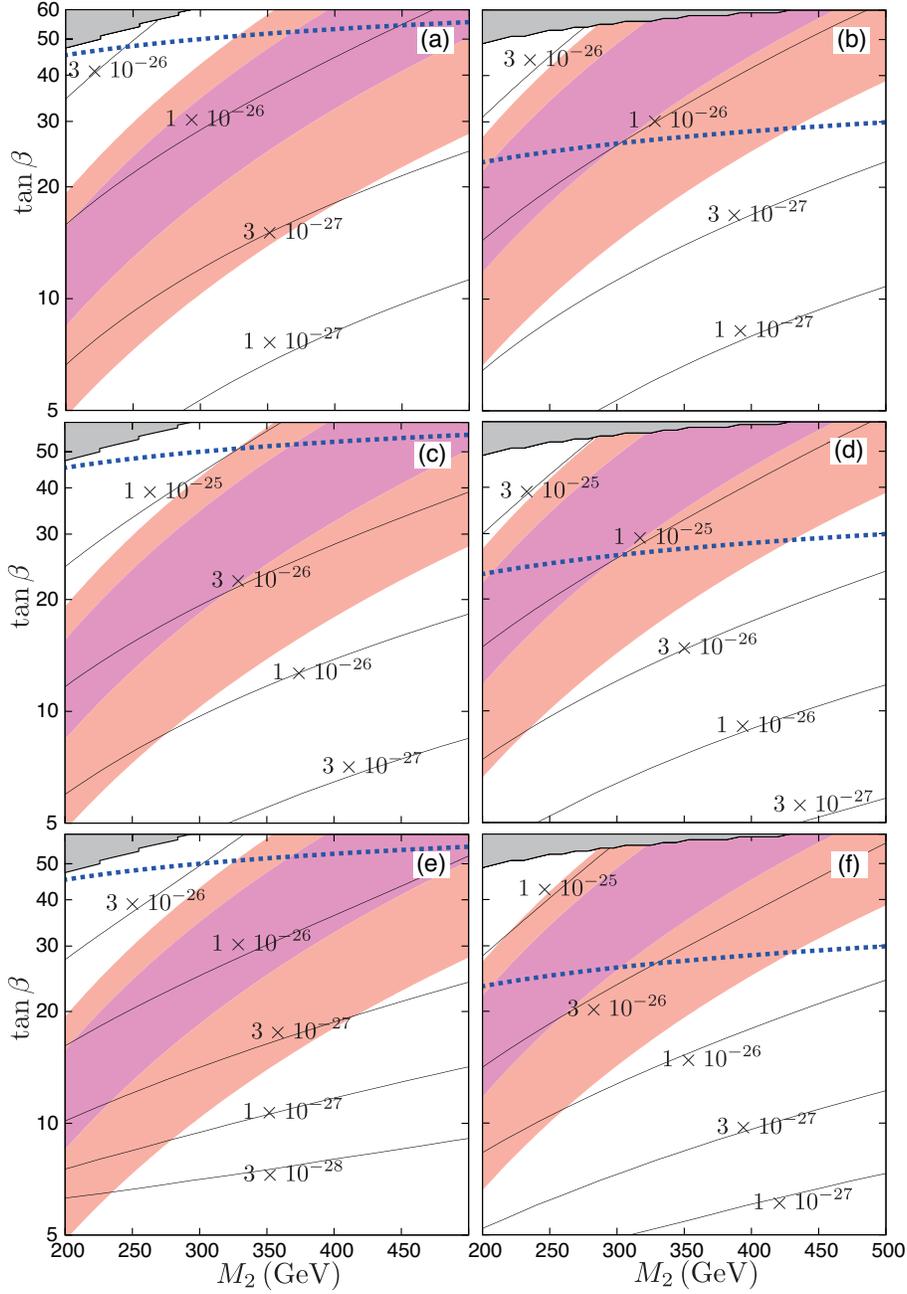}}
  \caption{\small Contours of EDMs of the electron ((a), (b)), the
    neutron ((c), (d)) and $\tilde{d}_q$ ((e), (f)) on $M_2$
    vs.\ $\tan\beta$ plane.  Here, we take $|\epsilon_{\bf
      24}|=10^{-2}$, and ${\rm arg}(\epsilon_{\bf 24})=\frac{\pi}{2}$,
    $M_{\rm mess}=10^{6}\ {\rm GeV}$ ((a), (c), (e)) and
    $10^{12}\ {\rm GeV}$ ((b), (d), (f)).  The sign of $\mu_H$
    is positive.  Blue dotted lines correspond to $B_\mu(M_{\rm
      mess})=0$.  In the pink (orange) regions, the muon MDM becomes
    consistent with the experimental value at $1\ \sigma$ ($2\
    \sigma$) level.  The gray regions are excluded due to
    un-successful electroweak symmetry breaking.  The 
    numbers in the figures
    are EDMs in units of $e\ {\rm cm}$ ((a) $-$ (d)) or 
    $\tilde{d}_q$ in units of ${\rm cm}$ ((e) and (f)).}
  \label{fig:edm24}
\end{figure}

The electron EDM gives a severe constraint on $\phi_A$. The
experimental bound on the electron EDM is given by
\cite{Nakamura:2010zzi}
\begin{eqnarray}
  d_e < 2.1 \times 10^{-27} e\ {\rm cm} ~~~ (95\% \, {\rm C.L.}). 
  \label{eq:bound_e_edm}
\end{eqnarray}
This bound should be compared to the SUSY contributions, which are
dominated by chargino diagrams.  The results are shown in panel (a)
and (b) of Fig.\ \ref{fig:edm24} on $M_2$ vs.\ $\tan\beta$ plane;
here, the normalization parameter $\Lambda$ is varied to give the
proper value of $M_2$, while $\tan\beta$ is calculated as a function
of the $B_\mu$ parameter at the messenger scale $B_\mu(M_{\rm mess})$.
In panels (a) and (b), the messenger scales are taken to be $M_{\rm
  mess}=10^6\ {\rm GeV}$ and $M_{\rm mess}=10^{12}\ {\rm GeV}$,
respectively.  One can see that the electron EDM is more enhanced as
$\tan\beta$ becomes larger, which is due to the fact that the
left-right mixing in the slepton sector is approximately proportional
to $\tan\beta$.

One important constraint on the MSSM parameter space may be from the
consideration of the anomalous magnetic dipole moment (MDM) of the
muon.  Recent detailed analysis shows that the deviation between the
experimental and theoretical values is at the $3.2\ \sigma$ level,
which may be due to the physics beyond the standard model
\cite{Cho:2011rk}.  Assuming that the deviation is from the SUSY
particles, the SUSY contribution to the muon MDM is \cite{Cho:2011rk}
\begin{eqnarray}
  a_\mu^{\rm (SUSY)} = (25.9 \pm 8.1)\times 10^{-10}.
  \label{amu(exp)}
\end{eqnarray}
In Fig.\ \ref{fig:edm24}, we also show the region where the muon MDM
becomes consistent with the above constraint.  One can easily see
that, in such a region, the electron EDM becomes one order of
magnitude larger than the experimental constraint if $|\epsilon_{\bf
  24}|=10^{-2}$ (as far as the phase is of $O(1)$).

Hadronic EDMs also give stringent constraints on the size and the
phase of $\epsilon_{\bf 24}$. Here we consider the neutron EDM and the
mercury EDM, induced by the EDMs and chromoelectric dipole moments
(CEDMs) of quarks.  The experimental bound on the neutron EDM is given
by \cite{Nakamura:2010zzi}
\begin{eqnarray}
  d_{n}< 2.9 \times 10^{-26} e\ {\rm cm} ~~~ (95\%\, {\rm C.L.} ),
\end{eqnarray}
which should be compared with theoretical estimation
\cite{Pospelov:1999ha, Pospelov:2000bw}\footnote
{The contributions from the strange quark are also discussed in the
  literature \cite{Hisano:2004tf}, and this may be as important as
  contributions from the up and down quarks. With inclusion of the
  strange quark contribution, the constraint from the neutron EDM can
  become as severe as the constraints from the mercury EDM and the
  electron EDM.}
\begin{eqnarray}
  d_n \simeq 1.4(d_d - 0.25d_u)+1.1e(d^c_d +0.5d^c_u), 
  \label{eq:neutron_EDM}
\end{eqnarray}
with $d_q$ and $d_q^c$ being the EDM and CEDM of a quark,
respectively.  Because the SUSY contributions to the down quark (C)EDM
is enhanced by $\tan\beta$, and also because $d_d$ and $d^c_d$ have
larger coefficient in Eq.(\ref{eq:neutron_EDM}), SUSY contribution to
$d_n$ is dominated by $d_d$ and $d_d^c$ in most of the cases.  We have
calculated $d_n$, and the results are shown in panels (c) and (d) of
Fig.\ \ref{fig:edm24} with $M_{\rm mess}=10^6\ {\rm GeV}$ and $M_{\rm
  mess}=10^{12}\ {\rm GeV}$, respectively. The parameter
$|\epsilon_{\bf 24}|=10^{-2}$ is marginally consistent for $M_{\rm
  mess}=10^6\ {\rm GeV}$, while it is excluded for $M_{\rm
  mess}=10^{12}\ {\rm GeV}$ in the parameter region consistent with
the muon MDM constraint \eqref{amu(exp)}.

The constraint from the mercury EDM is also stringent.  The
experimental bound is given by \cite{Griffith:2009zz}
\begin{eqnarray}
  |d_{Hg}| < 3.1 \times 10^{-29} e\ {\rm cm} ~~~ (95\%\, {\rm C.L.}) ,
\end{eqnarray}
which can be translated to the upper bound on the CEDMs of quarks as
\cite{Pospelov:2005pr, Ellis:2008zy}
\begin{eqnarray}
  \tilde{d}_{q} \equiv |d_d^c -d_u^c| \lesssim 
  4.4 \times 10^{-27} {\rm cm}. 
  \label{eq:dtq}
\end{eqnarray}
The contours of $\tilde{d}_q$ are shown in panels (e) and (f) with
$M_{\rm mess}=10^6 \ {\rm GeV}$ and $M_{\rm mess}=10^{12}\ {\rm GeV}$,
respectively.  The constraint from the mercury EDM is as stringent as
the electron EDM constraint.

In Fig.\ \ref{fig:edm24}, the contours of $B_\mu(M_{\rm mess})=0$ are
also shown, which correspond to vanishing $\phi_A$ in the absence of
the GUT relation breaking operator.  One can see that, adopting the
natural value of the $\epsilon_{\bf 24}$ parameter, the EDMs become too
large to be consistent with the experimental constraints if we adopt
\eqref{amu(exp)}.  Thus, if the breaking of the GUT relation is due to
an operator which is linear in $G_{\rm GUT}$ breaking fields, the SUSY
CP problem might remain even in models with vanishing $B_\mu$
parameter.

Next, we consider the case that $SU(5)_{\rm GUT}$ is broken by a field
in ${\bf 75}$ representation, as in the case of missing partner model
\cite{Masiero:1982fe, Grinstein:1982um};\footnote
{If the GUT symmetry is broken by a field in a real (but not adjoint)
  representation of $SU(5)_{\rm GUT}$, the leading order contribution
  of the GUT symmetry breaking is always $(M_{\rm GUT}/M_{\rm Pl})^2$
  as far as all the operators allowed by the $SU(5)_{\rm GUT}$
  symmetry are present.  Then, in those cases, the sizes of the EDMs
  are of the same order of the case with the GUT symmetry breaking by
  a ${\bf 75}$ representation. }
we parameterize the VEV as
\begin{eqnarray}
  \langle \Sigma^{ab}_{cd} \rangle
  &=& 
  \frac{v_{\bf 75}}{4\sqrt{6}}
  (\delta^a_c \delta^b_d - \delta^a_d \delta^b_c),
  \\
  \langle \Sigma \rangle^{a'b'}_{c'd'}
  &=& \frac{3v_{\bf 75}}{4\sqrt{6}}
  (\delta^{a'}_{c'} \delta^{b'}_{d'} - \delta^{a'}_{d'} \delta^{b'}_{c'}),
  \\
  \langle \Sigma \rangle^{ab'}_{cd'}
  &=& -\frac{v_{\bf 75}}{4\sqrt{6}}
  \delta^a_c \delta^{b'}_{d'},
\end{eqnarray}
where, in the above expressions, $a$, $b$, $\cdots$ run $1-3$, while
$a'$, $b'$, $\cdots$ run $4-5$.  (Then, the $X$- and $Y$-boson mass is
$m_{X,Y}=g_5 v_{\bf 75}$.)  In this case, the effect of the
$SU(5)_{\rm GUT}$ breaking in the superpotential should be second (or
higher) order in $v_{\bf 75}$; we consider the following
superpotential:
\begin{eqnarray}
  W &=& \lambda_0 S \bar{\Psi}^\alpha \Psi_\alpha
  + \frac{\lambda_1}{M_{\rm Pl}^2} 
  S \bar{\Psi}^\alpha \Sigma^{\gamma\delta}_{\alpha\epsilon}
  \Sigma^{\beta\epsilon}_{\gamma\delta} \Psi_\beta
  \nonumber \\ &=&
  \lambda_0 \left( 1 + \frac{1}{12}\epsilon_{\bf 75}^2 \right )
  S \bar{\psi}_d \psi_d 
  + \lambda_0 \left( 1 + \frac{1}{4}\epsilon_{\bf 75}^2 \right )
  S \bar{\psi}_l \psi_l + \cdots,
\end{eqnarray}
where
\begin{eqnarray}
  \epsilon_{\bf 75}
  = \frac{\lambda_1^{1/2} v_{\bf 75}}{\lambda_0^{1/2} M_{\rm Pl}}.
\end{eqnarray}
In this case, the gaugino masses are given by
\begin{eqnarray}
  M_A &=& \frac{g_A^2}{16\pi^2} 
  (1 + \kappa^{({\bf 75})}_A \epsilon_{\bf 75}^2)
  \Lambda,
\end{eqnarray}
with $\kappa^{({\bf 75})}_1=\frac{11}{60}$, $\kappa^{({\bf
    75})}_2=\frac{1}{4}$, $\kappa^{({\bf 75})}_3=\frac{1}{12}$.  

\begin{figure}
  \centerline{\epsfysize=0.8\textheight\epsfbox{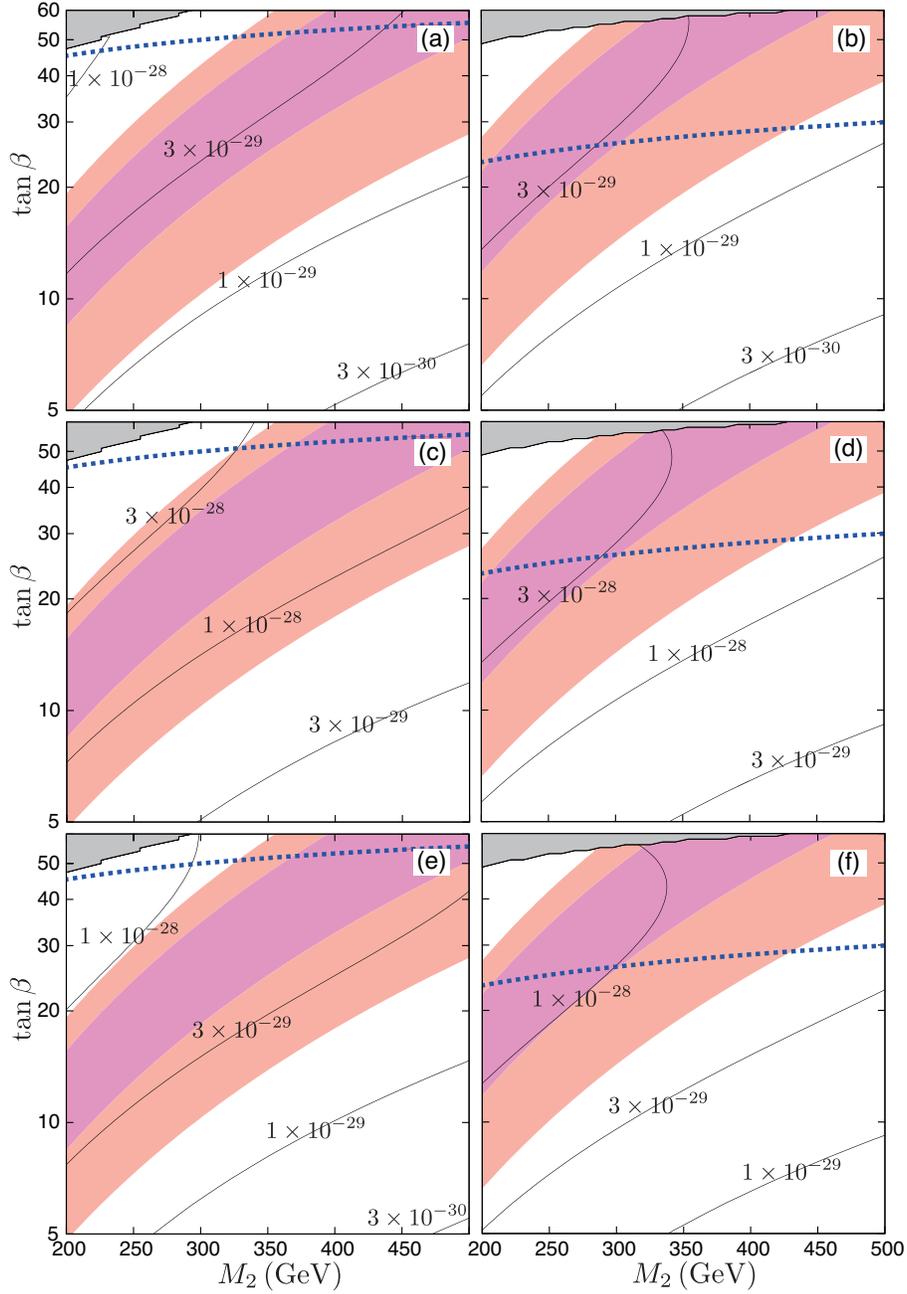}}
  \caption{\small Same as Fig.\ \ref{fig:edm24}, but with
    $|\epsilon_{\bf 75}|=10^{-2}$ and ${\rm arg}(\epsilon_{\bf
      75})=\frac{\pi}{4}$ (and $\epsilon_{\bf 24}=0$).}
  \label{fig:edm75}
\end{figure}

In Fig.\ \ref{fig:edm75}, we show the EDMs with $|\epsilon_{\bf
  75}|=10^{-2}$.  We take ${\rm arg}(\epsilon_{\bf
  75})=\frac{\pi}{4}$, which maximizes the EDMs. In panels (a), (c)
and (e), the messenger scale is taken to be $M_{\rm mess}=10^{6}\,{\rm
  GeV}$, while $M_{\rm mess}=10^{12}\,{\rm GeV}$ in (b), (d) and
(f). The electron EDM is shown in the panels (a) and (b), while neutron
EDM are shown in panels (c) and (d). In addition, the results of
$\tilde{d}_q$ are shown in panel, (e) and (f).  As one can expect, the
EDM constraints are less severe compared to the case with ${\bf 24}$
representation.  This indicates that the EDM constraints can be
avoided if operators linear in the $G_{\rm GUT}$ breaking field are
somehow forbidden.  However, such models are testable if the
experimental sensitivities on the EDMs are improved by two orders of
magnitude.
  
Before closing this section, we comment on another effect, i.e.,
renormalization group effect, which modifies the GUT relation among
the coupling constants.  The renormalization group equations of
$\lambda^{(Q)}_{ij}$ and $M^{(Q)}_{ij}$ are given by
\begin{eqnarray}
  \frac{d \lambda^{(Q)}_{ij}}{d \ln \mu} &=& 
  \gamma^{(\bar{Q})}_{ik} \lambda^{(Q)}_{kj} + 
  \gamma^{(Q)}_{jk} \lambda^{(Q)}_{ik} +
  \gamma^{(S)} \lambda^{(Q)}_{ij},
  \\
  \frac{d M^{(Q)}_{ij}}{d \ln \mu} &=& 
  \gamma^{(\bar{Q})}_{ik} M^{(Q)}_{kj} + 
  \gamma^{(Q)}_{jk} M^{(Q)}_{ik},
\end{eqnarray}
where $\gamma^{(X)}\equiv -\frac{1}{2} \frac{\partial \ln
  Z_X}{\partial \ln \mu}$ (with $Z^{(X)}$ being the wave-function
renormalization factor) is the anomalous dimension of the chiral
superfield $X$.  Then, the solutions of the above equations are given
by
\begin{eqnarray}
  \lambda^{(Q)}_{ij} (\mu) &=& 
  \zeta^{(\bar{Q})}_{ik} (\mu) \zeta^{(Q)}_{jl} (\mu) \zeta^{(S)} (\mu) 
  \lambda^{(Q)}_{kl} (M_{\rm GUT}),
  \\
  M^{(Q)}_{ij} (\mu) &=& 
  \zeta^{(\bar{Q})}_{ik} (\mu)  \zeta^{(Q)}_{jl}  (\mu)
  M^{(Q)}_{kl} (M_{\rm GUT}),
\end{eqnarray}
where
\begin{eqnarray}
  \zeta^{(Q)}_{ij} = \delta_{ij} + \sum_{n=1}^\infty
  \int_{\ln M_{\rm GUT}}^{\ln \mu} dt_1
  \int_{\ln M_{\rm GUT}}^{t_1} dt_2
  \cdots   \int_{\ln M_{\rm GUT}}^{t_{n-1}} dt_{n}
  \gamma^{(Q)}_{i k_1} (t_1)
  \gamma^{(Q)}_{k_1 k_2} (t_2) \cdots
  \gamma^{(Q)}_{k_{n-1} j} (t_n),
\end{eqnarray}
and
\begin{eqnarray}
  \zeta^{(S)} = \exp \left[ \int_{\ln M_{\rm GUT}}^{\ln \mu} dt
    \gamma^{(S)} (t) \right],
\end{eqnarray}
with $t=\ln\mu$.  A similar expression holds for
$\zeta^{(\bar{Q})}_{ik}$.  Then, for multiplets $Q$ and $Q'$
originating from the same multiplet of $G_{\rm GUT}$, the equality
$\Lambda^{(Q)}=\Lambda^{(Q')}$ is realized if
$\lambda^{(Q)}_{ij}=\lambda^{(Q')}_{ij}$ and
$M^{(Q)}_{ij}=M^{(Q')}_{ij}$ at the GUT scale.  Using the fact that
$\sum_Q c_A^{(Q)}$ become independent of the standard-model gauge
group (i.e., $A=SU(3)_{\rm C}$, $SU(2)_{\rm L}$, and $U(1)_{\rm Y}$)
as far as all the fields in the same multiplet of $G_{\rm GUT}$ are
contained in the summation, $m_A/g_A^2$ becomes universal and hence
there is no relative phase among gaugino masses.

\section{Effect of Supergravity}

Next, we consider the phases from the supergravity effect.  When the
standard model is supersymmetrized, naturally we consider local
supersymmetry, i.e., supergravity.  Then, supergravity effect may also
induce small but non-negligible CP violating phases.

Concerning the phases $\phi_A$, one should consider the effect on the
$B_\mu$ parameter.\footnote
{ $A$ parameters are also generated by the supergravity effect, and
  are also of the order of the gravitino mass.  The effect of the
  phases in the $A$ parameters are at most of the same order of the
  that in $B_\mu$ because the contribution of the $A$ parameter does
  not have the $\tan\beta$ enhancement. }
Naturally, such an effect is estimated to be of the order of the
gravitino mass $m_{3/2}$.  Importantly, in general, the supergravity
contribution to $B_\mu$ is complex, and its phase is not aligned to
those of gaugino masses.  Thus, this becomes a new source of CP
violation.

To study its consequence, we calculate the EDMs taking account of the
supergravity effects on the phase in $B_\mu$ parameter.  Here, we
parameterize $B_\mu$ as
\begin{eqnarray}
  B_\mu = B_\mu^{\rm (0)} + B_\mu^{\rm (SUGRA)},
\end{eqnarray}
where $B_\mu^{\rm (SUGRA)}$ is the supergravity contribution to the
$B_\mu$ parameter, which is expected to be of the order of the
gravitino mass.  Because we are interested in the gauge mediation
model, the gravitino mass is much smaller than the electroweak scale,
and hence $|B_\mu^{\rm (SUGRA)}|\ll |B_\mu|$.  As mentioned before, we
do not specify the source of the dominant contribution $B_\mu^{\rm
  (0)}$; we assume that $\phi_A\rightarrow 0$ as $B_\mu^{\rm
  (SUGRA)}\rightarrow 0$ to derive a conservative constraint.  In
addition, we adopt the usual GUT relation among the gaugino masses;
the $\epsilon_{\bf 24}$ and $\epsilon_{\bf 75}$ parameters used in the
previous section are set to be zero.

\begin{figure}
  \centerline{\epsfysize=0.8\textheight\epsfbox{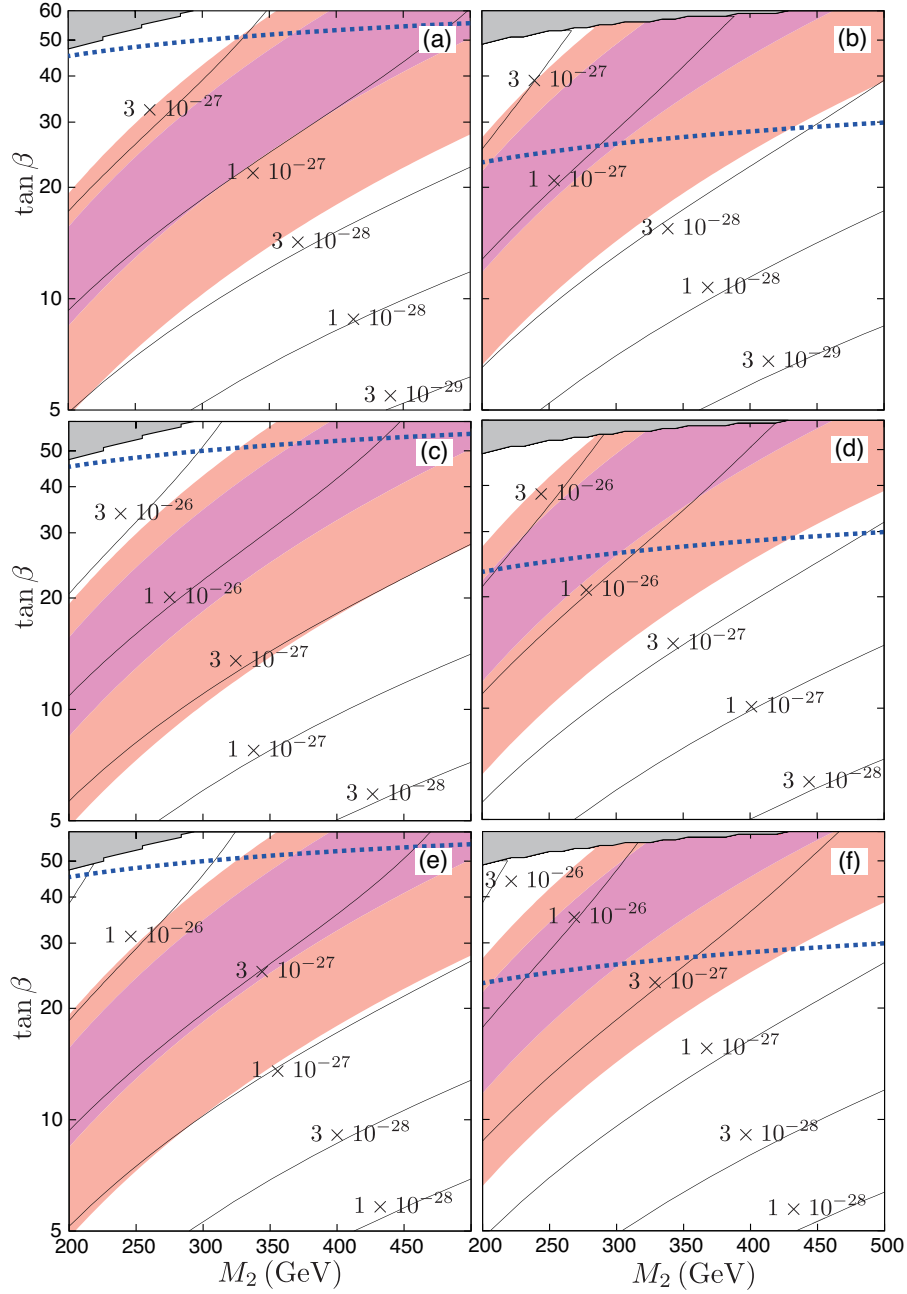}}
  \caption{\small Same as Fig.\ \ref{fig:edm24}, but with
    $|B_{\mu}^{\rm SUGRA}|=100\ {\rm MeV}$ and ${\rm arg}(B_{\mu}^{\rm
      SUGRA})=\frac{\pi}{2}$ (and $\epsilon_{\bf 24}=\epsilon_{\bf
      75}=0$).}
  \label{fig:sugra}
\end{figure}

We calculate EDMs taking $|B_\mu^{\rm (SUGRA)}|=100\ {\rm MeV}$ and
${\rm arg}(B_\mu^{\rm (SUGRA)})=\frac{\pi}{2}$.  With this choice, the
phase $\phi_A$ is of the order of $10^{-3}$.  The contours of constant
EDMs are shown in Fig.\ \ref{fig:sugra}.  One can see that, the
results are not sensitive to the messenger scale in the present set up
in which the messenger scale is not related to the gravtino mass.  In
the parameter region which is consistent with the constraint
\eqref{amu(exp)}, the electron and mercury EDMs become marginally
consistent with the experimental constraints when the gravitino mass is
$\sim 100\ {\rm MeV}$.  Notice that the supergravity contributions to
the off-diagonal elements of the sfermion mass matrices are typically
$\sim m_{3/2}^2$.  Thus, the CP violations may put severer upper bound
on the gravitino mass than the flavor violations.

In the present framework, the EDMs are proportional to ${\rm
  Im}(B_\mu^{\rm (SUGRA)})$ as far as $|B_\mu^{\rm (SUGRA)}|$ is much
smaller than $|B_\mu|$.  Because $|B_\mu^{\rm (SUGRA)}|$ is expected
to be of the order of the gravitino mass $m_{3/2}$, we can see that
gauge mediation models with $m_{3/2}\gtrsim 100\ {\rm MeV}$ require
some mechanism to suppress the phase in $B_\mu^{\rm (SUGRA)}$
parameter (and hence that in the gravitino mass) relative to the
gaugino masses if we take the anomaly in the muon MDM seriously.

\section{Summary}

In this Letter, we have studied the SUSY CP problem in gauge mediation
model.  We have paid particular attention to the effects of GUT
symmetry breaking and the supergravity effect.  Both of these effects
possibly induce too large CP violating phase in the MSSM parameters to
be consistent with the experimental constraints on EDMs.  It was
considered that the SUSY CP problem can be avoided in the gauge
mediation model in which $B_\mu$ parameter is not generated at the
messenger scale.  However, the CP violating phases discussed in this
letter cannot be eliminated even in such a model.

In particular, the effect of the GUT symmetry breaking spoils the GUT
relation among the gaugino masses if a mass term is introduced to the
vector-like messenger multiplets; consequently, the phases of the
gaugino masses may become different.  In the case that operators which
are linear in $G_{\rm GUT}$ breaking field are allowed, the EDMs are
likely to be too large in the parameter region where the SUSY
contribution to the muon MDM explains the discrepancy between the
experimental and standard-model values.  In other cases, the EDMs are
expected to be smaller than the current experimental bounds.  However,
even so, the effects may be observed if the experimental sensitivity
to the EDMs can be improved by two orders of magnitude.

\section*{Acknowledgements}
The authors thank R. Sato and K. Yonekura for useful discussion.  The
work of T.M. is supported by Grant-in-Aid for Scientific resear ch
from the Ministry of Education, Science, Sports, and Culture (MEXT),
Japan, No.\ 22540263 and No.\ 22244021, and the work of N.Y. is
supported by Grand-in-Aid for Scientific Research, No.22-7585 from
JSPS, Japan.

\end{document}